\documentclass[epjCONF,columns]{svjour} 
\usepackage{amsmath}
\usepackage{graphicx}

\usepackage[varg]{txfonts} 
\usepackage[latin1]{inputenc}
\usepackage{hyperref}
\usepackage{color}

\session-title{Hadron Collider Physics Symposium 2011}

\begin{document}

\title{Higgs Phenomenology of Minimal Universal Extra Dimensions}

\author{Genevi\`eve B\'elanger\inst{1} \fnmsep
    \thanks{\email{belanger@lapp.in2p3.fr}}
  \and Alexander Belyaev\inst{2} \fnmsep
    \thanks{\email{a.belyaev@soton.ac.uk}}
  \and Matthew Brown\inst{3} \fnmsep \thanks{\email{m.s.brown@soton.ac.uk}}
  \and Mitsuru Kakizaki\inst{4} \fnmsep
    \thanks{\email{kakizaki@sci.u-toyama.ac.jp}}
  \and Alexander Pukhov\inst{5} \fnmsep \thanks{\email{pukhov@lapp.in2p3.fr}}
}
  
\institute{
  LAPTH, Universit\'e de Savoie, CNRS, B.P.110, F-74941 Annecy-le-Vieux
    Cedex, France
  \and NExT Institute, School of Physics and Astronomy, University of
    Southampton, UK
  \and School of Physics and Astronomy, University of Southampton, UK
  \and Department of Physics, University of Toyama, 3190 Gofuku, Toyama
    930-8555, Japan
  \and Skobeltsyn Institute of Nuclear Physics, Moscow State
    University, Moscow 119992, Russia
}

\abstract{
The minimal model of Universal Extra Dimensions (MUED) is briefly reviewed. We
explain how the cross-sections for Higgs production via gluon fusion and decay
into photons are modified, relative the the Standard Model (SM) values, by KK
particles running in loops, leading to an enhancement of the $gg \to h \to
\gamma\gamma$ and $gg \to h \to W^+ W^-$ cross-sections. ATLAS and CMS
searches for the SM Higgs in these channels are reinterpreted in the context
of MUED and used to place new limits on the MUED parameter space. Only a small
region of between 1 and 3 GeV around $m_h = 125~\text{GeV}$ for
$500~\text{GeV}< R^{-1} < 1600~\text{GeV}$ remains open at the 95~\%
confidence level.
}
\maketitle

\section{Introduction}
\label{sec:intro}

The idea of extra dimensions of space is conceptually intriguing and provides
a rich framework for building models that go beyond the Standard Model (SM).
One class of extra-dimensional theories, proposed by Appelquist, Cheng and
Dobrescu \cite{PhysRevD.64.035002} in 2001, involves \emph{universal} extra
dimensions (UED). In such models all particles can propagate in all dimensions
(i.e. in the ``bulk''), and the extra dimensions are hidden by compactifying
them on a distance scale too small to be probed by our current experiments.

In this note we focus on the simplest possible UED model, Minimal UED (MUED),
which has one extra dimension compactified on a circle. The circle has
translational symmetry and so there is conserved momentum in the 5th dimension
which is discretised because of the periodic boundary conditions; one
therefore talks about conserved ``Kaluza-Klein'' (KK) number $n$. The extra
dimension is then further compactified onto an $\mathrm{S}^1/\mathbb{Z}_2$
orbifold (essentially a line segment with particular boundary conditions for
the fields ). This ``orbifolding'' is necessary to obtain chiral matter, and
it breaks the translational invariance so KK number is only conserved at tree
level. There is a remnant reflection symmetry of the orbifold about its
midpoint which leads to a KK ``parity'' $(-1)^n$ being conserved at all orders
in perturbation theory.

When MUED is expressed as a 4D effective theory, the different possible KK
numbers of a 5D particle manifest themselves as a tower of progressively
heavier separate 4D particles, one for each KK number. The zero mode particles
are identified with SM particles.

KK parity conservation makes MUED particularly hard to observe at the LHC
because it means that $n=1$ resonances (the lightest new particles predicted
by MUED) can only be pair produced and $n=2$ resonances are typically too
heavy to be produced on-shell at LHC energies. However, KK parity ensures that
the lightest $n=1$ KK particle (the ``LKP'' in analogy to the LSP in SUSY) is
stable; for much of the parameter space the LKP is neutral and so it provides
an excellent Dark Matter WIMP candidate. This is the main motivation for MUED.

Like all gauge theories in greater than four dimensions, MUED is
perturbatively non-renormalisable; the theory must be cut off at a momentum
scale $\Lambda$ above which the new physics of the UV completion is expected
to become noticeable. Discounting the mild cutoff dependence at LHC energy
scales, MUED has two parameters with values not completely constrained by
experiment: the Higgs mass $m_h$ and the inverse compactification
radius $R^{-1}$.

The $R^{-1}$ parameter is already bounded to be less than around 1600~GeV by
WMAP observations because it sets the scale of the Dark Matter (DM): if DM
were heavier it would lead to the Universe having a measurable positive
curvature. Also, $m_h$ is bounded from above by the requirement that DM be
neutral. These DM considerations are discussed in detail in
\cite{Belanger:2010yx} and shown in Fig.~\ref{fig:comblimits} of this note. 

The Higgs mass is required to be greater than 114.1~GeV by the LEP2 Higgs
searches and $R^{-1} \gtrsim 300~\text{GeV}$ in order not to conflict with
electroweak precision measurements \cite{Gogoladze:2006br}.

In this paper we place new constraints on the parameter space using data from
the latest ATLAS and CMS Higgs searches. This idea has been attempted before
\cite{Nishiwaki:2011gkV1}, but without taking into account the effect of MUED on
Higgs decay to two photons.\footnote
{
  Whilst we have been preparing these proceedings, Nishiwaki \emph{et al} have
  produced new work that includes KK suppression of the $h\to\gamma\gamma$
  vertex \cite{Nishiwaki:2011gk}.
}
This is explained more fully in the following
sections.

\section{Higgs production and decay} 
\label{sec:higgs_production_and_decay}

\begin{figure}[htbp]
  \begin{center}
    \includegraphics{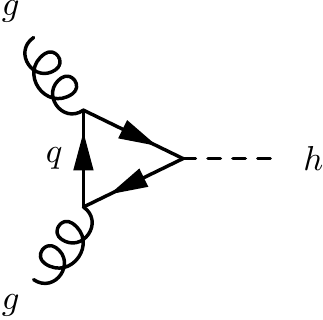}~\\
    \includegraphics{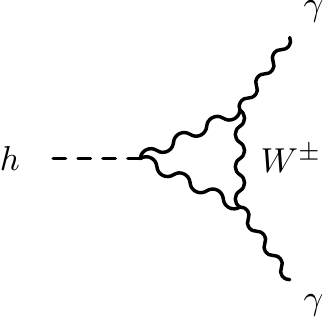}~~~~~~~~~~~~~~
    \includegraphics{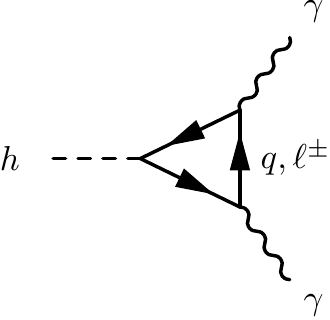} 
  \end{center}      
  \caption{Representative diagrams involved in Higgs production by gluon fusion
(top) and subsequent decay to photons (bottom).}
  \label{fig:diagrams}
\end{figure}

For low and intermediate $m_h$, the most important Higgs production process at
the LHC is gluon-gluon fusion. The lowest order Feynman diagram contributing
to this process is the one shown in Fig.~\ref{fig:diagrams} (top), with quarks
running in the triangle loop. For low $m_h$, Higgs decay to two photons is the
most important channel due to its small background. In Fig.~\ref{fig:diagrams}
(bottom), two leading order contributions are shown. There are actually many
other one-loop diagrams involving $W^{\pm}$ bosons that are not shown; for the
full details see \cite{belanger:2012}. But essentially there is tension
between the quark/lepton (dominated by the top quark) and the $W$
contributions to the decay, with the $W$ contribution being dominant in the
SM.

\begin{figure}[htbp]
  \begin{center}
	  \includegraphics{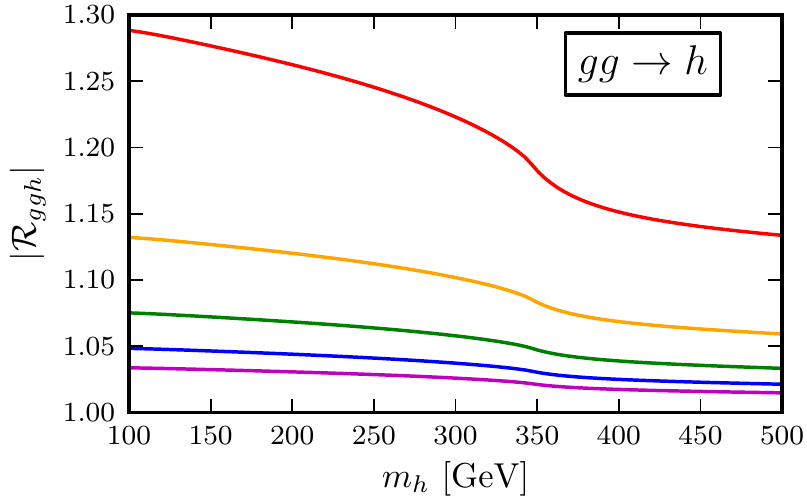} \\~\\
    \includegraphics{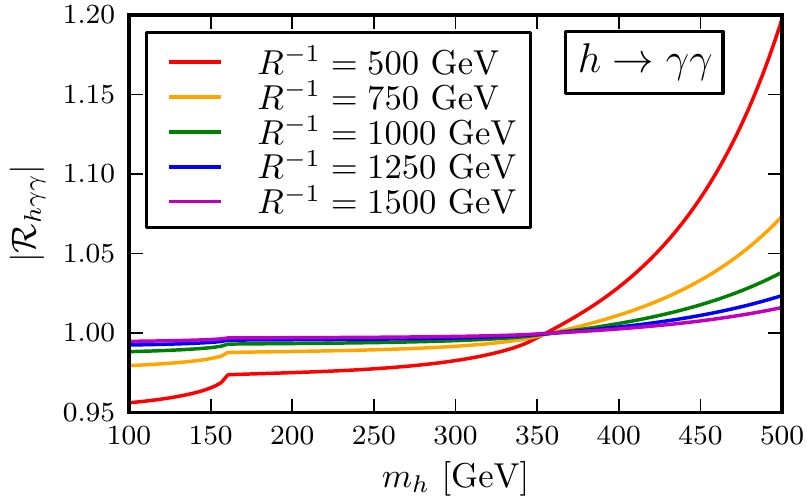} \\~\\ 
	  \includegraphics{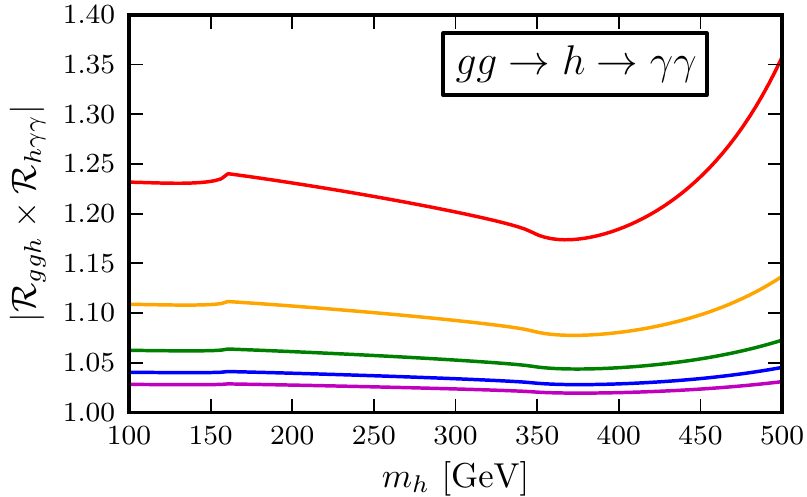}
  \end{center}
  \caption{Ratios of the scalar parts of matrix elements for Higgs production
  and decay, where $\mathcal{R} = \tilde{\mathcal{M}}_{\text{MUED}} / \tilde{\mathcal{M}}_{\text{SM}}$. For each graph, on the RHS, from top to bottom, the curves represent $R^{-1}$ values of 500, 750, 1000, 1250 and 1500~GeV.}
  \label{fig:ratios}
\end{figure}

In MUED, KK particles can also flow in the loops. These enhance each of the
diagrams shown. The rate of Higgs production is therefore increased relative
to the SM. The opposing fermion and $W$ contributions to the diphoton decay
are each enhanced, but the fermion enhancement is greater than the $W$ and so
the partial decay width to photons is \emph{decreased} for most of the
relevant $(m_h,R^{-1})$ parameter space. The matrix elements for production
and decay both take the form
\[
\mathcal{M} = \tilde{\mathcal{M}}\times\left(\frac{m_h^2}{2}g^{\mu\nu} - p^\nu
q^\mu \right) \epsilon_\mu \epsilon_\nu,
\]
where the $\epsilon$ four-vectors are gluon or photon polarisations. We have
approximated external particles as being on shell. The scalar parts
$\tilde{\mathcal{M}}$ of the matrix elements for $g,g\to h$ and $h\to
\gamma,\gamma$ are plotted in Fig.~\ref{fig:ratios} (top and middle) as
multiples of the SM values for various values of $m_h$ and $R^{-1}$.

We used our own implementation of the MUED model in the \texttt{LanHEP} and
\texttt{CalcHEP} software packages in order to calculate these matrix elements
and, later, cross-sections. Unlike other implementations, ours includes the
effects of radiative corrections to KK particle masses at one-loop because
these corrections play a vital role due to the highly-degenerate tree-level
MUED mass spectrum \cite{Cheng:2002iz}. Our model also includes two-loop SM corrections to the
$ggh$ and $h\gamma\gamma$ vertices that can be as large as 50~\% of the
leading order values, although these cancel in the ratios plotted in
Fig.~\ref{fig:ratios}.

Overall, the enhancement of the Higgs production amplitude is greater than the
suppression of the Higgs decay to two photons and so the MUED cross-section
for $gg \to h \to \gamma\gamma$ is always enhanced relative to the Standard
Model's. This is shown the bottom graph in Fig.~\ref{fig:ratios}. This means
that our model is more sensitive to experimentally-determined Higgs mass
limits that the SM and this sensitivity can be used to constrain the parameter
space further.

In addition to $gg \to h \to \gamma\gamma$, we also looked at $gg \to h \to W^+ W^-$, which is particularly important in the intermediate Higgs mass range. The gluon fusion part of the process is enhanced as before, but decay to two $W$s can proceed via a tree-level diagram so KK particles have no leading order effect on the decay part.

\section{Limits on parameter space} 
\label{sec:limits_on_parameter_space} We looked at the latest ATLAS and CMS SM
Higgs searches and reinterpreted the analyses for MUED. The results of the
searches are expressed in terms of $\mu \equiv \sigma_{95\%} /
\sigma_{\text{SM}}$, where $\sigma_{95\%}$ is the cross-section for a
particular Higgs production and decay process that has currently been ruled
out at the 95~\% confidence level, and $\sigma_{\text{SM}}$ is the SM
cross-section for that same process.

One can place limits on MUED by calculating $\mu_{\text{MUED}} =
\sigma_{\text{MUED}} / \sigma_{\text{SM}}$ for different values of $(m_h,
R^{-1})$ and seeing whether it is larger than the existing limit. We used the
latest limits shown in Fig.~3 of \cite{ATLAS-CONF-2011-163} for ATLAS and
Fig.~6 (top) of \cite{CMS-PAS-HIG-11-032} for CMS. We combined the limits from the two experiments statistically for each of
channels of interest (diphoton and $W^+ W^-$) using
\[
\mu_{\text{comb}} = \left(\frac{1}{\mu_\text{ATLAS}^2} +
\frac{1}{\mu_\text{CMS}^2}\right)^{-\frac{1}{2}}.
\]
This does not take into account systematic errors but it does give a good
estimate of the combination \emph{in lieu} of the official combination from
ATLAS and CMS.

\begin{figure}[htbp]
  \begin{center}
  	\includegraphics{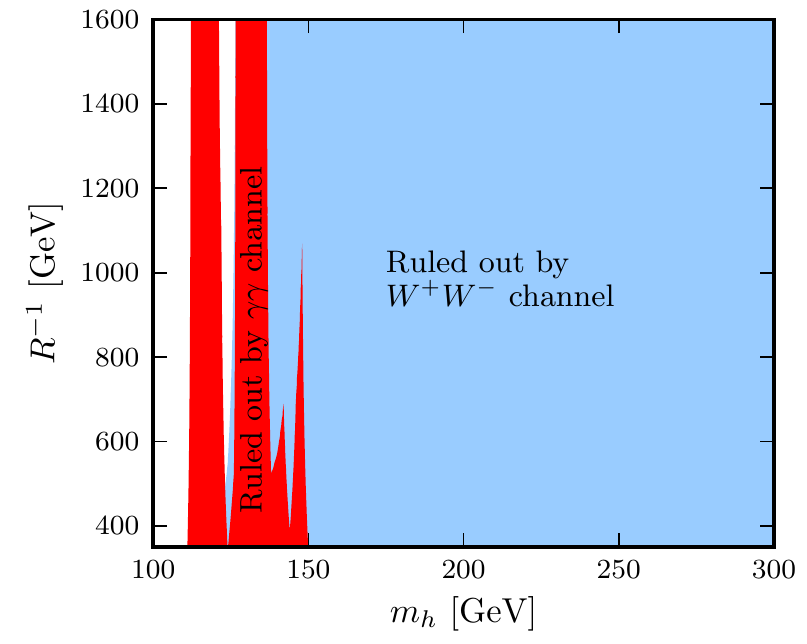}
  \end{center}
  \caption{Regions of MUED parameter space ruled out at the 95~\% confidence
level by combination of ATLAS and CMS Higgs searches using the diphoton (red)
and $W^+W^-$ (blue) channels.}
  \label{fig:mylimits}
\end{figure}

The parameter space ruled out by the Higgs search data is shown in
Fig.~\ref{fig:mylimits}. All of the parameter space for $m_h > 111~\text{GeV}$
is ruled out except for a small region around 125~GeV -- this is due to the
excess of events observed by ATLAS and CMS recently in this region.

\begin{figure}[htbp]
  	\includegraphics[width=\columnwidth]{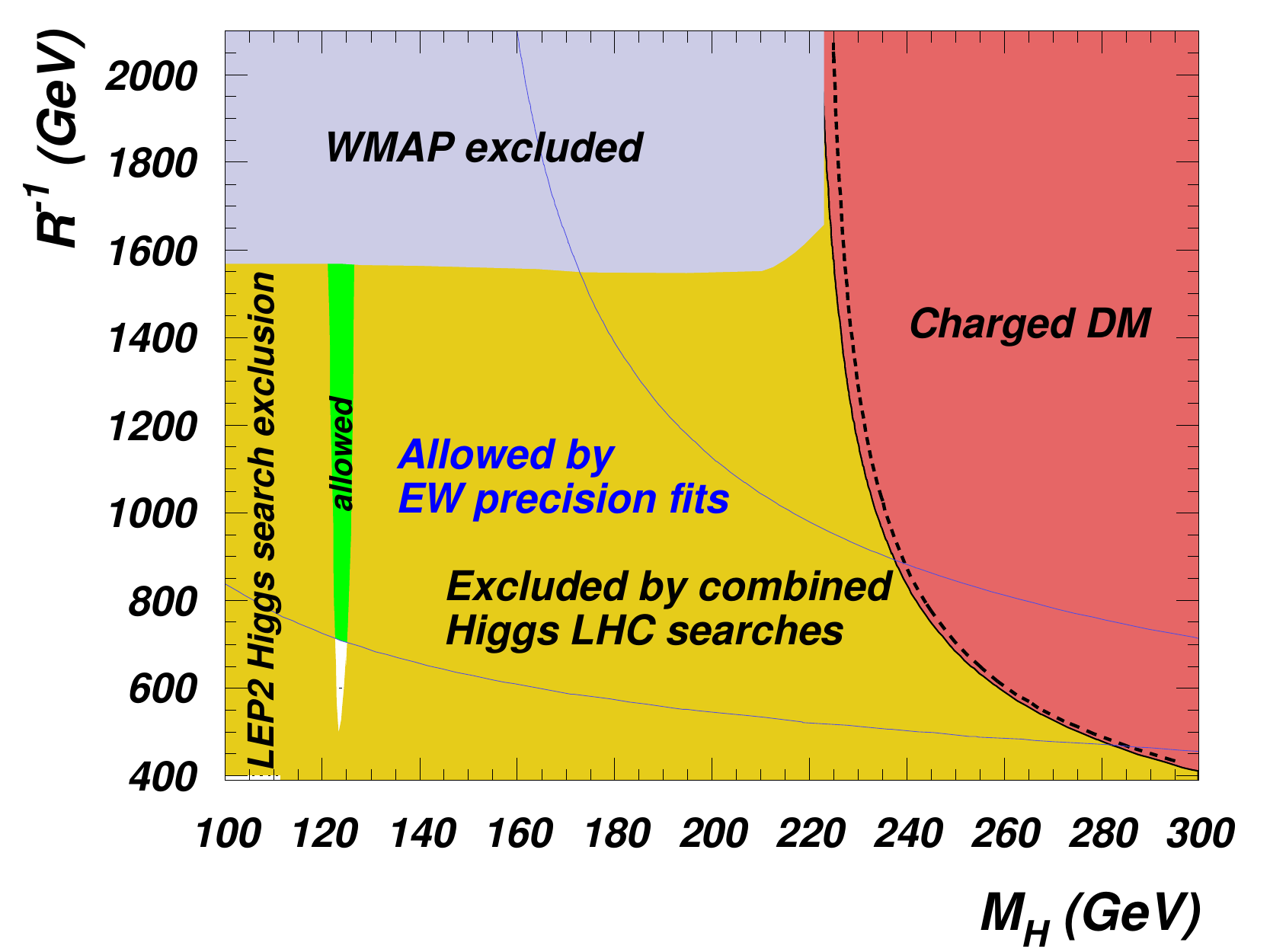}
  \caption{Constraints on MUED parameter space. Purple and pink show regions
ruled out by DM considerations. Gold denotes the region excluded by the Higgs
search analysis presented here and also the existing LEP2 limit.
Points between the blue hyperboloids agree with LEP EW precision fits to
a 95~\% CL.}
  \label{fig:comblimits}
\end{figure}

For completeness, in Fig.~\ref{fig:comblimits} we show the limits on the MUED
parameter space from the Higgs analysis presented here together with existing
limits from other constraints. In addition to those constraints described in
Section~\ref{sec:intro}, electroweak precision fits from LEP prefer a Higgs
with a mass in the window delineated on the graph by the two blue
hyperboloids.

What is left is a very narrow region of parameter space with $m_h$ around
125~GeV and $700~\text{GeV} < R^{-1} < 1600~\text{GeV}$.

\section{Conclusions}

We have used the latest ATLAS and CMS Higgs search data to constrain the
parameter space of MUED. We have improved on an earlier analysis by including
the effects of the KK modes on the Higgs decay to two photons and by also
including the radiative corrections to the KK masses. Full details will be
given in \cite{belanger:2012}.

We eagerly await the official limit combination from ATLAS and CMS, although
details of the combination will become moot (for our purposes!) if the 125~GeV
excess goes away when extra data is collected in 2012. If this happens, we
will be able to rule out MUED completely at the 3$\sigma$ limit by the end of
the year. If the excess remains and we discover the Higgs there, this should
allow us to make a prediction as to the value of $R^{-1}$.


\begin{thebibliography}{1}
\bibitem{PhysRevD.64.035002}
T.~Appelquist, H.-C. Cheng, and B.~A. Dobrescu, \href{http://dx.doi.org/10.1103/PhysRevD.64.035002}{{\em Phys.
  Rev. D} {\bfseries 64} no.~3, (2001) 035002}

\bibitem{Belanger:2010yx}
G.~Belanger, M.~Kakizaki, and A.~Pukhov,
\href{http://arxiv.org/abs/1012.2577}{{\ttfamily arXiv:1012.2577 [hep-ph]}}

\bibitem{Gogoladze:2006br}
I.~Gogoladze and C.~Macesanu,
  \href{http://dx.doi.org/10.1103/PhysRevD.74.093012}{{\em Phys. Rev.}
  {\bfseries D74} (2006) 093012},
\href{http://arxiv.org/abs/hep-ph/0605207}{{\ttfamily arXiv:hep-ph/0605207}}

\bibitem{Nishiwaki:2011gkV1}
K.~Nishiwaki, K.-y. Oda, N.~Okuda, and R.~Watanabe,
\href{http://arxiv.org/abs/1108.1764v1}{{\ttfamily arXiv:1108.1764v1 [hep-ph]}}

\bibitem{belanger:2012}
G.~Belanger, A.~Belyaev, M.~Brown, M.~Kakizaki, and A.~Pukhov, ``In
  preparation''

\bibitem{Cheng:2002iz}
H.-C. Cheng, K.~T. Matchev, and M.~Schmaltz, 
  \href{http://dx.doi.org/10.1103/PhysRevD.66.036005}{{\em Phys. Rev.}
  {\bfseries D66} (2002) 036005},
\href{http://arxiv.org/abs/hep-ph/0204342}{{\ttfamily arXiv:hep-ph/0204342}}

\bibitem{ATLAS-CONF-2011-163}
{The ATLAS Collaboration}, Tech. Rep. ATLAS-CONF-2011-163, (December 2011)

\bibitem{CMS-PAS-HIG-11-032}
{The CMS Collaboration}, CMS-PAS-HIG-11-032, (December 2011)

\bibitem{Nishiwaki:2011gk}
K.~Nishiwaki, K.-y. Oda, N.~Okuda, and R.~Watanabe, 
\newblock {\em Phys.Lett.} \newblock {\bf B707} (2012) 506--511



\end{thebibliography}
\small
\providecommand{\href}[2]{#2}\begingroup\raggedright\endgroup

\vskip-30pt
\section*{Acknowledgements}
\vskip-8pt
{\small
AB and MB thank Royal Society for partial financial support. AB also
does so with the NExT Institute and SEPnet.
}
\end{document}